\documentclass[12pt,aps,prb,preprint]{revtex4}   

\usepackage{amsmath}    
\usepackage{graphicx}   

\begin{document}

\title{ac-driven Brownian motors: \\ a Fokker-Planck treatment}
\author{S. Denisov}
\email{sergey.denisov@physik.uni-augsburg.de}
\author{P. H\"{a}nggi}
\affiliation{Institut f\"ur Physik, Universit\"at  Augsburg,
       Universit\"atsstra{\ss}e 1, D-86135 Augsburg, Germany}
\author{J. L. Mateos}
\affiliation{Instituto de F\'{\i}sica,
Universidad Nacional Aut\'onoma de M\'exico, \\
Apartado Postal 20-364, 01000 M\'exico, D.F., M\'exico}

\date{\today}

\begin{abstract}
We consider a primary model of ac-driven Brownian motors, i.e., a
classical particle placed in a spatial-time periodic potential and
coupled to a heat bath. The effects of fluctuations and dissipations
are studied by a time-dependent Fokker-Planck equation. The approach
allows us to map the original stochastic problem onto a system of
ordinary linear algebraic equations. The solution of the system
provides complete information about ratchet transport, avoiding such
disadvantages of direct stochastic calculations  as long transients
and large statistical fluctuations. The Fokker-Planck approach to
dynamical ratchets is instructive and opens the space for further
generalizations.
\end{abstract}

\maketitle

\section{Introduction}

When one starts to pan for gold, he/she places a gold bearing sand
(or gravel) into a pan and then agitates it in a circular motion,
back and forth, left and right. Finally, the sand goes away with the
water and gold grains remain in the bottom of the pan. However, if
one simply inclines the pan  then all the content will go away.

This is the illustration of the basic mechanism which lies behind
the ratchet idea: in a driven system without a preferable direction
of motion, transport velocities depend on characteristics of movable
objects -- charges, spins, masses, sizes, etc \cite{RMP_BM,HH,
HMN,Reim}. The intensive development of this idea during the last
decade brought new tools for a smart control of transport in
different systems, ranging from mechanical engines \cite{Slava} to
quantum devices \cite{Linke}.

The ratchet setup demands three key ingredients which are (i)
nonlinearity, (ii) asymmetry (spatial and/or temporal), and (iii)
fluctuating input force of zero mean. The nonlinearity, in a sense
of a nonlinear system response to an external force, is necessary
since, otherwise, the system will produce a zero-mean output from a
zero-mean input. The asymmetry is needed  to violate the left/right
symmetry of the response  because  transport in a fully symmetrical
system is unbiased. The zero-mean fluctuating force should break
thermodynamical equilibrium, which forbids appearance of a directed
transport  due to the Second Law of Thermodynamics
\cite{Fein,Parrondo}.

The basic model of dynamical ratchets is an one-dimensional
classical particle in a periodic potential exposed to an ac-field
\cite{Flach1, Ketz, Hanggi2, Mateos}. While the symmetry analysis of
microscopic equations of motion \cite{Flach1} enables one to
formulate necessary conditions for the directed transport appearance
\cite{Ren}, both the current sign and strength depend on dynamical
mechanisms. At the deterministic limit, when noise is absent, the
evolution of a damped particle is governed by attractors, regular
(limit cycles) or chaotic ones \cite{Ott}. Transport properties are
encoded in the characteristics of attractors, so when there is only
one attractor in phase space, the dc-current is equal to a mean
velocity of the attractor\cite{Mateos}.

The situation changes drastically when noise starts to contribute to
the dynamics. During the evolution, particle jumps out of an
attractor, evolves outside of the attractor vicinity, lands back
into the attractor, etc. In other words, the particle explores the
\textit{whole phase space} and the dynamics of the particle can not
be described in terms of the attractor properties only. The
situation becomes even more complicated  when several attractors
coexist in phase space \cite{Mateos1}.

The standard approach, based on the direct Langevin simulations of
the microscopic equations of motion \cite{Langevin},  demands a very
long time in order to overcome  transient effects and to produce the
sufficient self-averaging over a phase space. The goal of this paper
is to show that all these problems can be tackled by using the
time-dependent Fokker-Planck equation (FPE) \cite{Risken, Jung}. Our
approach allows us to reduce the stochastic problem to a system of
ordinary linear algebraic equations, which can be solved by using
standard numerical routines. The solution of the system of equations
provides a full information on transport properties of the system.

The outline of this article is as follows. In Sec.II we introduce
the model and set up the problem within the Fokker-Planck frame.
Then, we formulate the symmetries which need to be violated in order
to get a nonzero dc-current. Section III discusses the method of the
solution of ac-driven  Fokker-Planck equations. In Sec. IV, we
illustrate our approach with the so-called tilting  ratchets
\cite{Reim}. Sec. V contains some conclusions and perspectives.

\section{The model}

The one-dimensional classical dynamics of a particle (e.g. a cold
atom placed into an optical lattice, or a colloidal microsphere in a
magnetic bubble lattice \cite{HH}), of mass $m$ and friction
coefficient $\gamma$, exposed to an ac-driven periodic potential and
coupled to a heat bath, is described by the equation:
\begin{eqnarray}
m \ddot{x} + \gamma \dot{x} = g(x,t) + \xi(t), ~~ g(x,t) = -
\partial U(x,t)/\partial x, \label{particle}
\end{eqnarray}
where $U(x,t)$ is a potential and the force $g(x,t)$ is time and
space periodic,
\begin{eqnarray}
g(x+L,t)=g(x,t+T)=g(x,t). \label{force}
\end{eqnarray}
The noise is modeled by a $\delta$-correlated Gaussian white noise,
$\langle \xi(t) \rangle = 0, ~~ \langle \xi(t) \xi(t') \rangle = 2
\gamma D \delta(t-t')$, where the noise intensity is $D=kT$.

The state of the system (\ref{particle}) can be represented as a
point in the three-dimensional phase space, $(x, v, t)$, for $m \neq
0$ (underdamped regime) and in   the two-dimensional phase space,
$(x,t)$, when $m = 0$ (overdamped limit) \cite{Ott}. The stationary
asymptotic current is equal to
\begin{equation}
J=\lim_{t \rightarrow \infty} J(t)=\lim_{t \rightarrow \infty}
\frac{x(t)}{t}. \label{current}
\end{equation}

The statistical description of the system at $m=1$ is provided by
the the Fokker-Planck equation \cite{Risken, Jung}
\begin{equation}
\{\partial_t+ \frac{\partial}{\partial x} v -
\frac{\partial}{\partial v}   [\gamma v -g(x,t)]- \gamma D
\frac{\partial^{2}}{\partial v^{2}}\}P(x,v,t)=0, \label{fpe_under}
\end{equation}
where $v=\dot x$. The respective FPE for the overdamped limit, when
inertia is neglible, $m=0$, reads \cite{Risken, Jung}
\begin{equation}
\gamma {\dot{P}}(x,t)=-[\frac{\partial}{\partial x} g(x,t)- D
\frac{\partial^{2}}{\partial x^{2}}]P(x,t)~. \label{fpe_over}
\end{equation}

The overdamped limit is the appropriate description for
microdynamics at low Reynolds number, when a particle moves in an
extremely viscous media \cite{Reynolds}.

The Fokker-Planck equations, (\ref{fpe_under}),(\ref{fpe_over}), are
linear, dissipative, and preserve the norm, $\int P dx dv$ for Eq.
(\ref{fpe_under}) and $\int P dx$ for Eq.(\ref{fpe_over})
\cite{Jung}. In addition, the equations possess discrete time and
space translation symmetries, so that the operations $x \rightarrow
x + L$ and $t \rightarrow t + T$ leave the equations invariant. For
a given boundary condition and a fixed norm, any initial
distribution, $P(...,0)$, will converge to a single time-periodic
attractor solution, $\tilde{P}(...,t)=\tilde{P}(...,t+T)$. What are
the correct spatially boundary condition for the ratchet problem
given by Eqs.(\ref{particle}-\ref{current})? It has been shown that
the current can be calculated with the spatial-periodic solution of
the type $\tilde{P}(x,...)=\tilde{P}(x+L,...)$ \cite{Reim}.

The dc-components of the directed current (\ref{current}) in terms
of the spatially-periodic attractor solution $\tilde{P}$ are given
by \cite{Reim}:

\begin{eqnarray}
&&J=\langle \int_{-\infty}^{\infty}v \cdot \hat{P}(x,v,t)\cdot dv
\rangle_{T,L},~~m=1~,
\label{cur_under}\\
&&J={\gamma}^{-1}\langle g(x,t) \cdot \hat{P}(x,t)\rangle_{T,
L},~~m=0~,\label{cur_over}
\end{eqnarray}

where $\langle ... \rangle_{T}=\int_{0}^{T}... dt$ and $\langle ...
\rangle_{L}=\int_{0}^{L}... dx$. Without loss of generality, we set
$L=2\pi$.

Let us assume that Eq.(\ref{fpe_under}) (or Eq.(\ref{fpe_over})) is
invariant under some transformation of the variables $x$ and $t$,
which does not affect the boundary conditions. Then the unique
solution $\tilde{P}$ is also invariant under the transformation. The
strategy is now to identify symmetry operations which invert the
sign of the current $J$ (\ref{cur_under}) (or (\ref{cur_over})) and,
at the same time, leave the corresponding FPE invariant. If such a
transformation exists, the dc-current $J$ will strictly vanish
\cite{Den}. Sign changes of the current can be obtained by either
inverting the spatial coordinate $x$, or time $t$ (plus simultaneous
velocity inversion, $v \rightarrow -v$, for the underdamped case).

Below we list all such transformations together with the
requirements the force $g(x,t)$  has to fulfill \cite{Flach1, Den}:
\begin{eqnarray}
{\widehat S}_1: x \to - x + x', ~t \to t+t';\; ~~~~~ {\widehat S}_1
(g) \to -g\;,
~~~~~~~~~~~~~~~~~~\label{S1}\\
{\widehat S}_2: x \to x + x',~t \to -t+t';~~~~~{\widehat S}_2 (g)
\to g\;,~~~~~~~(\mbox{if }\gamma=0).~\label{S2}
\end{eqnarray}

Here $x'$ and $t'$ depend on the particular shape of $g(x,t)$.

There is an additional symmetry for the overdamped limit
\cite{Super},
\begin{eqnarray}
{\widehat S}_3: x \to x + x',~t \to -t+t';~~~{\widehat S}_3 (g) \to
-g\;,~~~~~~~(\mbox{if }m=0).~\label{S2}
\end{eqnarray}

which does not follow from the FPE (\ref{fpe_over}). This is not a
symmetry of the equation (due to the last r. h. s. term in
Eq.(\ref{fpe_over}), so-called diffusive term \cite{Risken}). It has
been shown, nevertheless, that the symmetry ${\widehat S}_3$
corresponds to a certain property of the asymptotic attractor
solution $\tilde{P}(x,t)$, such that the current (\ref{cur_over})
exactly vanishes when $\widehat{S}_3$ holds \cite{Den}.

By a proper choice of $g(x,t)$ all relevant symmetries can be
broken, and then one can  expect the appearance of a non-zero dc
current $J$ (\ref{cur_under}-\ref{cur_over}).

\section{Method of solution}

It is impossible to solve the FPE's (\ref{fpe_under}-
\ref{fpe_over}) analytically.  Therefore we deal with the equations
by using the Fourier expansion (over the x and t) for the overdamped
case (\ref{fpe_over}),
\begin{eqnarray}
P(x,t)=\frac{1}{\sqrt{2\pi T}}\sum^{N,K}_{n,k=-N,-K}P_{nk} \cdot
e^{i(nx+k \omega t)}, \label{exp_over}
\end{eqnarray}
plus the matrix continued fraction technique \cite{Risken} for the
underdamped regime (\ref{fpe_under}),
\begin{eqnarray}
P(x,v,t)=\frac{\psi_{0}(v)}{\sqrt{2\pi T}}\sum^{N,K}_{n,k=-N,-K}
\sum^{S}_{s=0} P_{n k s} \cdot e^{i(nx+k \omega t)} \psi_{s}(v),
\label{exp_under}
\end{eqnarray}
where $\psi_{s}(v)$ is the Hermite function of $s$-order
\cite{Hermite}. Both the expansions (\ref{exp_over}-\ref{exp_under})
are truncated, so we can control the convergence and the precision
of the solution $\tilde{P}$ by variation of the parameters $N$, $K$
and $S$.

By inserting the expansions (\ref{exp_over}-\ref{exp_under}) into
the corresponding FPEs (\ref{fpe_over}-\ref{fpe_under}), we obtain
the following systems of linear algebraic equations,
\begin{eqnarray}
(i k \omega +D n^{2}) P_{lk} +i n \sum_{l,q} g_{l q}
P_{(n-l)(k-q)}=0, \label{sys_over}
\end{eqnarray}
for the overdamped limit, and
\begin{eqnarray}
(i k \omega+s \gamma D)P_{nks}+ i \sqrt{D} n (\sqrt{s}P_{n k (s-1)}
+ \sqrt{s+1} P_{n k (s+1)})- \sqrt{\frac{s}{D}}\sum_{l,q} g_{l q}
P_{(n-l)(k-q) (s-1)}=0, \nonumber
\\ \label{sys_under}
\end{eqnarray}
for the underdamped regime. Here $g_{nk}$ is the Fourier element of
the force function, $g(x,t)=\sum_{n,k} g_{nk}e^{i(nx+k \omega t)}$.
We note that the equation for the dc-element, $P_{00}$ in
Eq.(\ref{fpe_over}) ($P_{000}$ in Eq.(\ref{fpe_under})), should be
replaced by the equation $P_{00}=1$ ($P_{000}=1$). This corresponds
to the normalization condition, $\int_{0}^{2\pi}P(x,t)dx = 1$
($\int_{0}^{2\pi}\int_{-\infty}^{\infty} P(x,v, t)dx dv = 1$).

The definition for the current takes the form
\begin{eqnarray}
J= \sqrt{\frac{D}{LT}}\cdot P_{0 0 1}, \label{curR_under}
\end{eqnarray}
for the underdamped case, and
\begin{eqnarray}
J= \frac{1}{\sqrt{LT}}\sum_{n, k} g_{n k}\cdot P_{-n -k},
\label{curR_over}
\end{eqnarray}
for the overdamped limit.

In order to use the standard routine we have to transform the
original variables, $P_{nk}$ (overdamped limit) and $P_{nks}$
(underdamped regime), into the single-index variable
$\mathcal{P}_z$, $z=1,...,Z$. There is the one-to-one
transformation,
\begin{eqnarray}
\{n,k\} \rightarrow z=1+(n+N)(2K+1)+ k+K \label{trans1_over} \\
Z=(2N+1)(2K+1),\label{trans2_over}
\end{eqnarray}
for the overdamped limit, and
\begin{eqnarray}
\{n,k,s\} \rightarrow z=1+s(2N+1)(2K+1)+(n+N)(2K+1)+k+K
\label{trans1_under}
\\
Z=(2N+1)(2K+1)(S+1), \label{trans2_under}
\end{eqnarray}
for the underdamped regime, which transforms the original two- and
three-dimensional matrices, $P_{nk}$ and $P_{nks}$, into the
vector-column $\mathcal{P}_z$.

The corresponding system of equations for $\mathcal{P}_z$ can be
solved by using the standard numerical routines. For our
calculations, we have used the  routine F07ANF from Linear Algebra
Package (LAPACK) written in Fortran 77 \cite{Routine}.

\section{Example: tilting ratchets}

There is a plenty of choices for a driving potential $U(x,t)$
\cite{Reim}. Below we consider the case of a particle placed onto a
stationary periodic potential and driven by an alternating tilting
force \cite{Flach1,Hanggi2, Mateos, Den},
\begin{eqnarray}
g(x,t)=-V'(x)+E(t). \label{force_tilting}
\end{eqnarray}

\begin{center}
\begin{figure}[t]
\begin{tabular}{cc}
\includegraphics[width=0.6\linewidth,angle=0]{Fig1a.eps}\\
\includegraphics[width=0.6\linewidth,angle=0]{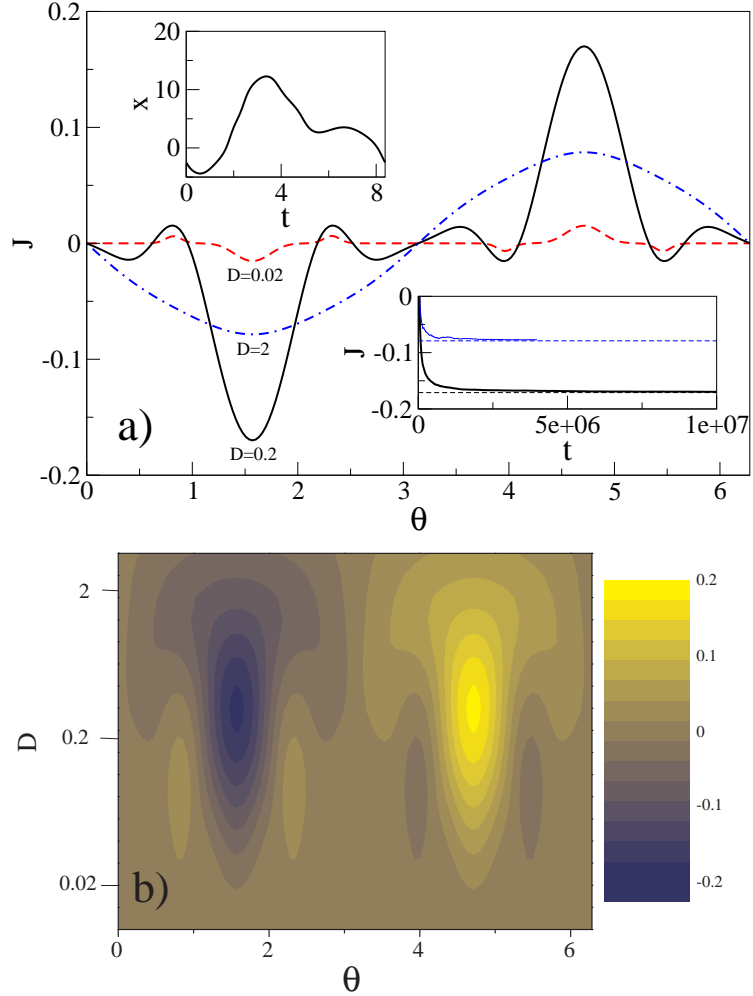}
\end{tabular}
\label{Figure1} \caption{ (color online)(a) Dependence of the
current $J$ (\ref{curR_over}) on $\theta$, for the system
(\ref{sys_over}) ($N=K=60$), at different temperatures, $D=kT$. The
parameter are $\gamma=1$, $V_{0}=2$, $E_1=4.6$, $E_2=-6$, and
$\omega=0.75$. Insets: the attractor of the corresponding
deterministic system (\ref{particle}) for $\theta=-\pi/2$(top left);
the dependence of the running current $J(t)=x(t)/t$ on $t$ at the
point $\theta=-\pi/2$, for $D=0.2$ (thick line) and $D=2$ (dashed
dotted line). Dashed lines correspond to the current values $J$
(\ref{curR_over}) (bottom right); (b) More detailed dependence: the
current $J$ (\ref{curR_over}) as a function of $\theta$ and $D$.}
\label{Fig:fig1}
\end{figure}
\end{center}

With the simple potential $V(x)=V_{0}(1-\cos(x))$, the two-frequency
driving,
\begin{equation}
E(t)=E_{1}\sin (\omega t)+E_{2}\sin (2\omega t +\theta),
\label{eq:driv}
\end{equation}
ensures that  for $E_1,E_{2} \neq 0$ $\hat{S}_1$ is always violated.
The symmetry $\hat{S}_2$ is broken for $\theta \neq 0, \pm \pi$. In
addition, $\hat{S}_3$ does not hold at $\theta \neq \pm\pi/2$
\cite{Den}.

We start from the overdamped limit, $m=0$ and $\gamma=1$. For a
given set of parameters, $V_0=2$, $E_1=4.6$, $E_2=-6$, and $\omega =
0.75$, there is only one limit cycle in the phase space of the
system (inset in Fig.1). The limit cycle is nontransporting, i. e.
particle returns to the initial position after one period of
external driving $T$, for the whole range $\theta \in [-\pi, \pi]$.
Therefore, there is no dc-transport in the deterministic limit,
$D=0$.

\begin{center}
\begin{figure}[t]
\begin{tabular}{cc}
\includegraphics[width=0.5\textwidth,angle=0]{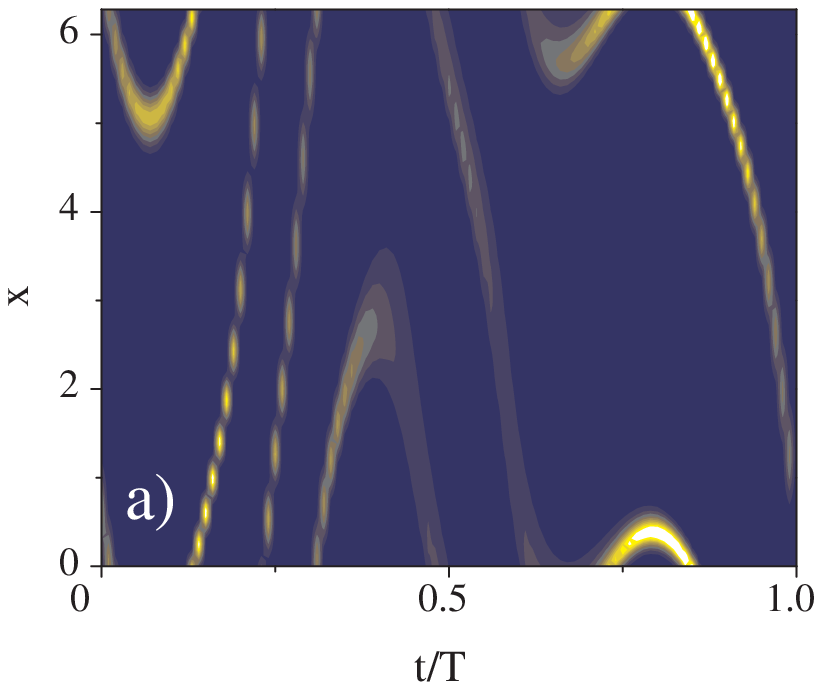}
\includegraphics[width=0.465\textwidth,angle=0]{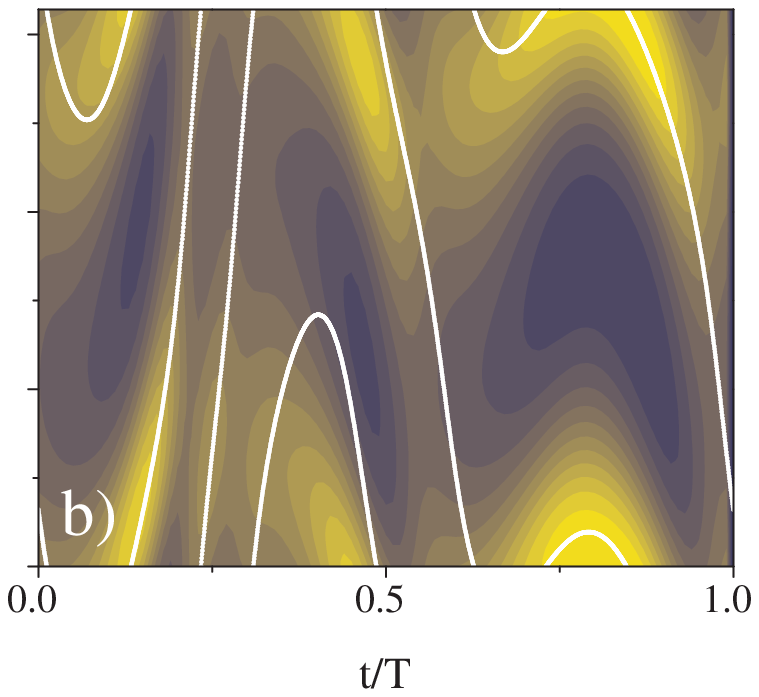}
\end{tabular}
\label{Figure1} \caption{(color online) The attractor solution
$\tilde{P}(x,t)$ of the FPE (\ref{fpe_over}),  for $D=0.02$ (a) and
$D=2$ (b). The white line in panel (b) marks the corresponding
deterministic attractor. Here $\theta=-\pi/2$, while the other
parameters are the same as in Fig.1.} \label{Fig:fig2}
\end{figure}
\end{center}

When noise enters  the game, it changes the dynamics: particle can
leave the attractor and this leads to a finite current appearance
(Fig.1). Therefore, thermal fluctuations play a constructive role
here by providing a way for the manifestation of the asymmetry
hidden in the ac-driving. Since the current depends nonmonotonously
on the noise intensity $D$, there is some kind of the stochastic
resonance effect \cite{st_resonance1, st_resonance2}. For a weak
noise, the dynamics is still localized near the attractor (Fig. 1a,
dashed line), so that the current is faint. In the opposite
high-temperature limit, when the noise starts to dominate the
dynamics, the particle dynamics is "smeared" over the phase space.
The relevant space-temporal correlations are suppressed by the noise
and the current tends to decrease.  There is a resonance temperature
(near $D \approx 0.27$, Fig.1b), where the dc-current can be
resonantly enhanced. Here the directed flow of particles is
analogous to the flow of information in the case of stochastic
resonance \cite{st_resonance1, st_resonance2}.

\begin{center}
\begin{figure}[t]
\includegraphics[width=0.6\linewidth,angle=0]{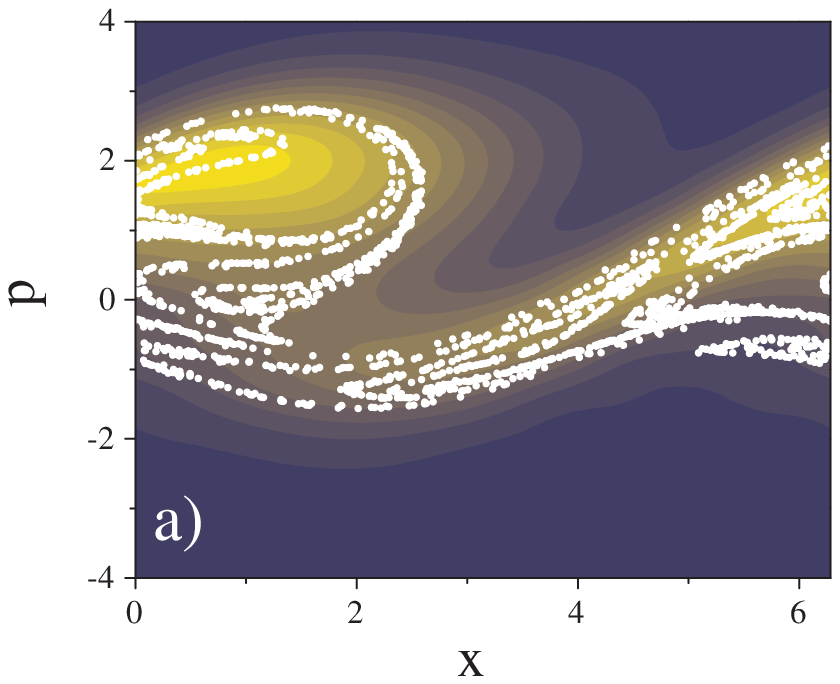}
\label{Figure1} \caption{(color online) The stroboscopic
representation of the attractor solution $\tilde{P}(x,v,0)$
(\ref{exp_under}) of the Eq. (\ref{sys_under}) ($N=K=35, S=25$),
with the Poincare section of the corresponding deterministic
attractor (\ref{particle}) (white dots). The parameters are $m=1$,
$V_{0}=1$, $E_1=E_2=1.2$, $\omega=1$, $\gamma = 0.1$, and
$\theta=0$.} \label{Fig:fig3}
\end{figure}
\end{center}

Note, that the noise variation can not violate the basic symmetries:
current is always absent at the point $\theta=0$ and $\theta=\pm
\pi$. We have corroborate our analysis by the direct Langevin
simulation (inset on Fig. 1) and found a perfect agreement with the
numerical solutions of the corresponding FPE.

The overdamped limit, however, is not suitable for the description
of all realistic situations. For example, for the modeling of
Josephson ratchets it is necessary to take into account the inertia
term, $m \ddot{x}$ \cite{Ustinov}. The underdamped regime without
periodic driving has been studied for a long time. To the best of
our knowledge, there is only one paper where FPE for an underdamped
system with ac-driving has been solved numerically \cite{Hanggi3}.

\begin{center}
\begin{figure}[t]
\includegraphics[width=0.9\linewidth,angle=0]{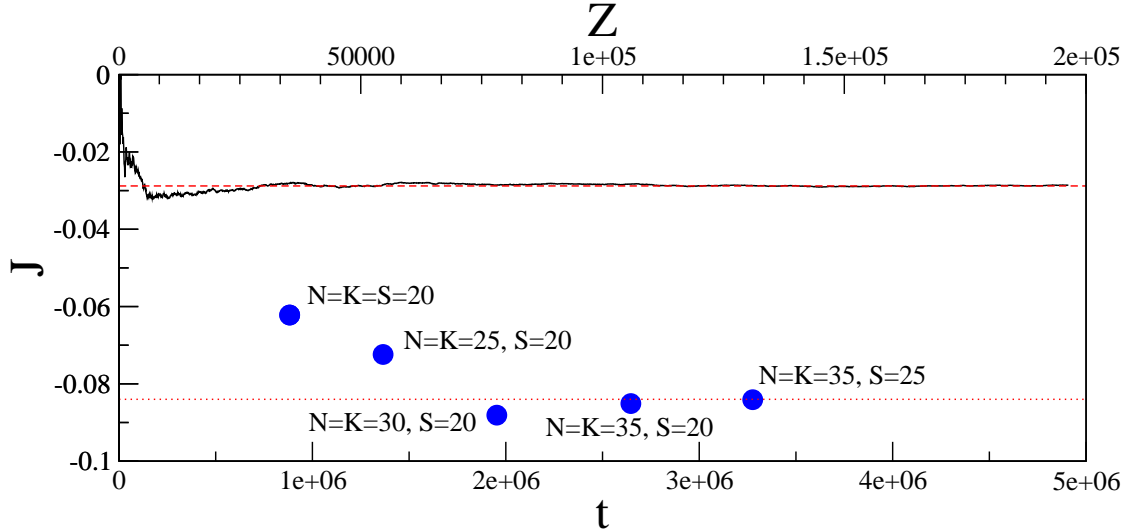}
\label{Figure1} \caption{(color online) The running current
$J(t)=x(t)/t$ vs $t$ for deterministic system (\ref{particle}) and
the dependence of the current $J$ (\ref{curR_under}) on the number
of basic states $Z$ (\ref{trans2_under}). The other parameters are
the same as in Fig.3.} \label{Fig:fig4}
\end{figure}
\end{center}

The results for $m=1$, $V_{0}=1$, $E_1=E_2=1.2$, $\omega = 1$,
$\gamma = 0.1$, and $\theta=0$ are shown in Fig.3. There is a single
chaotic attractor for the deterministic limit, $D=0$. In Fig.3a the
Poincare section of the attractor (the position of the particle in
the phase space, $[x(t),v(t)]$, taken at stroboscopical times
$t_n=nT$ \cite{Ott}), is depicted. Since
 the relevant symmetry, $S_1$, is violated by the presence of the second harmonics
$E_2$, the attractor generates a finite dc-current, $J \approx
-0.0288$ (Fig.4).

Although the attractor solution of the corresponding FPE for $D=0.1$
clearly resembles its deterministic counterpart (Fig.3), it produces
a much stronger current, $J \approx -0.083$, than in the
deterministic limit. Here, again, thermal fluctuations play a
constructive role in the process of dc-current rectification.

The convergence of the quantity (\ref{curR_under}) with increasing
of the total number of basic states, $Z$ (\ref{trans2_under}), is
relatively slow (Fig.4). We have to take into account $71$ spatial
and $71$ temporal harmonics in order to estimate accurately the
ratchet current. This fact means that the fast temporal modes
(chaotic dynamics on a short time scale, $t_c \ll T$) and  short
spatial modes ($x_c \ll L$),  contribute essentially to the overall
ratchet current.

\section{Conclusions}

We have shown that the complex issue  of ac-driven ratchets in a
thermal environment  can be resolved by using the Fokker-Planck
equation. Our approach maps the original problem onto a set of
ordinary algebraic linear equations, which then can be solved  by a
standard numerical routine.

The approach opens a new perspectives for further developments. For
example, it allows a generalization for the case of colored noise
with exponential correlation function \cite{Risken}. The
corresponding one-dimensional "non-thermal" process can be embedded
into a two-dimensional "thermal" dynamics by introducing the
additional dynamical variable, $\varepsilon(t)$, which should
replace the noise term, $\xi(t)$, in Eq.(\ref{particle})
\cite{colored}. The next prominent perspective is the quantum limit
of underdamped ac-driving ratchets. The master equation for the
corresponding Wigner function can be represented as a FPE
(\ref{sys_under}) with an additional coupling between elements
 \cite{ext_quantum}. Finally, the absolute
negative mobility in ac-driven inertial systems \cite{negative,
alatriste} is another possible target for our method.

\begin{acknowledgments}
This work has been supported by the DFG-grant HA1517/31-1.
\end{acknowledgments}


\begin{thebibliography}{500}
\bibitem{Hermite} We do not specify the Hermite functions since the Eqs.(14-15) provide
the full and self-consistent description of the problem. Function
definitions can be find in Ref.[17].
\end{thebibliography}
\end{document}